\documentclass{naturemod}
\usepackage{graphicx}

\title{Quantum interference between two single photons \\ emitted by
independently trapped atoms}

\author{J. Beugnon$^1$, M. P. A. Jones$^1$, J. Dingjan$^1$, B.
Darqui\'{e}$^1$,  G. Messin$^1$, A. Browaeys$^1$ \& P. Grangier$^1$}

\begin{document}

\maketitle
\begin{affiliations}
\item Laboratoire Charles Fabry de l'Institut d'Optique (UMR
8501), \\ B\^{a}timent 503, Centre Universitaire,
 91403 Orsay cedex, France .
\end{affiliations}

{\bf When two indistinguishable single photons are fed into the
two input ports of a beam splitter, the photons will coalesce and
leave together from the same output port. This is a quantum
interference effect, which occurs because the two possible paths
where the photons leave in different output ports interfere
destructively. This effect was first observed in parametric
downconversion by Hong, Ou and Mandel~\cite{Hong_Ou_Mandel87}, and
then with single photons produced one after the other by the same
quantum emitter~\cite{Santori02,Abram05,Rempe04,Zumbusch05}. With
the recent development of quantum information, a lot of attention
has been devoted to this coalescence effect as a resource for
quantum data processing using linear optics
techniques~\cite{KLM,Dowling,Pan,Riedmatten,Beige05,Barrett05}. To
ensure the scalability of schemes based on these ideas, it is
crucial that indistinguishable photons are emitted by a collection
of synchronized, but otherwise independent sources.  In this
paper, we demonstrate the quantum interference of two single
photons emitted by two independently trapped single atoms,
bridging the gap towards the simultaneous emission of many
indistinguishable single photons by different emitters. Our data
analysis shows that the coalescence observed is mostly limited by
the wavefront matching of the light emitted by the two atoms, and
to a lesser extent by the motion of each atom in its own trap.}

A basic requirement for most quantum computing schemes is the
implementation of two-qubit quantum gates ~\cite{Zoller}. If the
qubits are encoded in single photons, the gate can be obtained by
using an interference effect between the photons, followed by a
measurement-induced state projection~\cite{KLM}. One may also use
qubits encoded in solid-state systems such as quantum
dots~\cite{Gerard01}, or in atomic systems such as
ions~\cite{Wineland95} or neutral atoms~\cite{Meschede04}. One way
to entangle the atomic qubits without direct interaction, and thus
realise quantum gates, is to use them as single photon sources, so
that the emitted photons are entangled with the internal states of
the emitters. The interference of two photons emitted by such
sources projects the state of the two atoms into an entangled
state~\cite{Simon03}. Many protocols based on this conditional
entanglement have been proposed~\cite{Beige05,Barrett05}, and
experimental work is under way to implement them~\cite{Monroe04}.
The photons involved in such schemes do not need to be initially
entangled, and can even be emitted by different
sources~\cite{Riedmatten}, but they need to be indistinguishable.
However, it is generally not easy to have several (possibly many)
independent sources emitting indistinguishable photons. With
quantum dots in microcavities~\cite{Abram05,Santori02}, the
dispersion in frequency associated with differences in fabrication
is usually much too large for the photons to be emitted at the
same frequency. With atoms in cavities~\cite{Rempe04}, each
emitter is by itself a complicated experiment, and cannot be
easily multiplied. In this paper we address this problem by using
two single atoms in two neighbouring traps emitting in free space,
and we demonstrate that these atoms do emit indistinguishable
photons. This scheme can easily be scaled to arrays of
traps~\cite{Bergamini}.

Our experiment uses two single rubidium 87 atoms, confined in
separate optical dipole traps. These traps are formed in the focal
plane of the same high-numerical aperture lens, and loaded from a
cloud of cold atoms in an optical molasses~\cite{Schlosser01}. The
two traps, each of which has a waist of 1\,$\mu$m, are separated
by a distance of 6\,$\mu$m. To obtain triggered single photon
emission from the two atoms, we simultaneously excite them with a
high efficiency ($> 95\%$) using a $\sigma^+$-polarised pulsed
laser beam which drives the $F=2, \; m_F =2$ to $F'=3, \; m_F' =3$
closed transition~\cite{Dingjan06}. The quantization axis is
defined by a magnetic field of several gauss. Both atoms are
excited by the same short (less than 4 ns) $\pi$-pulse. Each one
then spontaneously emits a single photon~\cite{Darquie}, with a
lifetime of 26~ns. The photons are collected by the same objective
lens that is used to focus the dipole trap
beams~\cite{Schlosser01}, and detected using a pair of avalanche
photodiodes. Between the objective and the photodiodes, an optical
setup composed of two half-wave plates and two polarizing beam
splitter cubes (HWP1,2 and PBS1,2) is inserted in the beam path
(see Figure~1). It can be configured either as a 50/50 beam
splitter which mixes the light from the two atoms on each
detector, or as a beam separator which sends the light from each
atom to only one of the detectors. The avalanche photodiodes are
connected to a high-resolution counting card in a start-stop
configuration. This allows us to measure the number of coincident
photodetections as a function of the delay between the arrival
times of the two photons on the photodiodes, with a resolution of
about 1.2\,ns. In the 50/50 beam splitter configuration, the
detectors cannot distinguish which atom has emitted a photon, and
we expect to observe the coalescence effect. In the beam separator
configuration, each avalanche photodiode only monitors the light
emitted by one of the two atoms, and coincidence counts can only
be due to  independent emissions by both atoms.

The measurements are performed by repeating the following
procedure. First, we detect the simultaneous presence of one atom
in each trap in real time by measuring their fluorescence from the
molasses light used to load the traps. We then trigger a sequence
that alternates a burst of 575 pulsed excitations, lasting
115\,$\mu$s, with a 885\,$\mu$s cooling period using the molasses
light. This alternation is repeated 15 times before stopping and
recapturing a new pair of atoms. During the excitation periods,
the $\pi$-pulses irradiate both atoms every 200\,ns, and the
counting card accumulates the number of double detections produced
by the two avalanche photodiodes. This sequence maximizes the
number of single photons that we can obtain before the two atoms
are heated out of the trap~\cite{Darquie}. After the 15 bursts of
pulsed excitations, we measured a probability of 65\% to keep each
atom in its trap. At the end of the sequence, we switch the
molasses back on and wait on average about 300 ms until we detect
two atoms again. Two histograms are accumulated for the same
number of repetitions: one in the 50/50 beam splitter
configuration, and one in the beam separator configuration.

The two histograms are shown in Figure~2, without background
subtraction. Both histograms consist of a series of peaks separated by
200\,ns, the repetition period of the pulsed laser. The width of the
peaks is determined by the 26\,ns lifetime of the excited state. In the
beam separator configuration, all peaks are identical, and always
correspond to a double detection with one photon coming from each atom.
Hence, their height gives a natural calibration of the experiment. A
histogram in the 50/50 beam splitter configuration can be normalised by
dividing by the height of the peaks in its corresponding histogram
measured in the beam separator configuration. The normalized signals
that are obtained are then independent of collection efficiency,
detection efficiency and experiment duration, and allow histograms taken
under different conditions to be compared. In the 50/50 beam splitter
configuration, the peak at zero delay is clearly much smaller than the
other peaks. As each atom is a very good source of single
photons~\cite{Darquie}, this peak also consists only of events where both
atoms have emitted a photon. In contrast, the other peaks consist of
events where two photons are successively emitted, either by the same
atom, or by both atoms. Since the peaks at non-zero delay are almost the
same in both configurations, we can deduce that almost all registered
counts are due to events where both atoms were present.

In the case of perfect coalescence, the peak at zero delay in the
50/50 beam splitter configuration would be absent: as the two
photons leave via the same port, there can be no coincidences. We
attribute the residual peak that we observe in Figure~2 to an
imperfect overlap of the spatial modes of the two photons, which
then do not interfere. To experimentally illustrate this effect,
we varied the overlap between the two modes in a controlled way by
translating one beam relative to the other (translation of the cut
mirror CM, see figure~1). Figure~3 shows the normalized height $R$
of the residual peak at zero delay, as a function of the
separation between the two images. For a given spatial overlap $K$
between the amplitude of the two modes, the ratio $R$ is expected
to be $(1-K^2)/2$ (see Methods section). The solid line is a
gaussian fit based on the experimental value of the beam size in
the image plane, and considering the maximal wavefront overlap
$K_{\rm max}$ as an ajustable parameter.  The agreement with the
coincidence data is very good, which confirms the crucial role of
good mode matching of the two beams in our experiment. We obtain
from the fit the maximum wavefront overlap $K_{\rm max}=0.78\pm
0.03$. This imperfect overlap is consistent with the errors we
measure on the beam positions and waists.

Finally, we analyse the structure of the time spectrum around zero
delay. The small peak at zero delay from Figure~2, is displayed on
a larger scale in Figure~4. The dashed line correponds to a model
where the wavepackets of the two photons are identical and arrive
at the same time on the beamsplitter, but with imperfect spatial
overlap of the two beams. This curve does not correctly reproduce
the experimental data: the experimental dots seem to sit on a
slightly wider curve. Due to their finite temperature, the atoms
move in the trapping potential and experience a range of
light-shifts. This changes their internal energy, and thus
modifies the frequency of the emitted photon. For two photons at
different frequencies, the correlation signal would exhibit a beat
note as already seen by Legero {\it et al.}~\cite{Rempe04}. If the
two photons now have a distribution of frequencies, the
correlation signal consists of a beat note averaged out over this
distribution. This gives rise to a slightly broader structure for
the signal, which is well fitted by the solid line predicted by
our simple model (see details in the Methods section).

By fitting the experimental data shown in Figure~4 with our model,
we extract the overlap of the spatial modes $K=0.7\pm 0.05$ and
the temperature of the atoms $T=180\pm 20$ $\mu$K. In a separate
experiment, we measured the temperature of the atoms in the dipole
trap, which is initially close to $120\pm 10\,\mu$K. Each pulse
followed by the spontaneous emission of the photon increases the
energy of the atom by one recoil. We calculate that after the
first 115\,$\mu$sec of pulsed excitations the temperature rises
by 60\,$\mu$K, in good agreement with the temperature obtained
from the fit above. We also checked experimentally that each
cooling period resets the temperature of the atom to its initial
value.  A comparison of the fit with the dashed curve, which
corresponds to zero temperature, confirms that at present the
imperfect interference is mainly due to the imperfect optical
wavefront matching, and not to the motion of the atoms in the
traps.

In conclusion, we have experimentally demonstrated the coalescence
of two photons emitted  by two independent trapped atoms. The
contrast of the interference is limited by the overlap (in free
space) of the spatial modes of the fluorescence light emitted by
the two atoms. By coupling the light from each of the atoms into
identical single-mode optical fibres, this overlap could be
greatly improved, though this may be at the cost of a reduced
counting rate. The shape of the residual signal around zero delay
is well explained by a broadening due to the finite temperature of
the atoms in the trap. Better wavefront overlap and further cooling of the atoms will improve
the overall quality of this interference and will make this system
suitable as a resource for entangling two atoms.

\begin{addendum}
\item[Supplementary Information] is linked to the online version
of the paper at www.nature.com/nature.
 \item We acknowledge support from the European
Union through the Integrated Project ``SCALA". J. Dingjan was
funded by Research Training Network ``CONQUEST". M. Jones was
supported by a Marie Curie fellowship.
 \item[Competing Interests] The authors declare that they have no
competing financial interests.
 \item[Correspondence] Correspondence and requests for materials
should be addressed to A. Browaeys~(email:
antoine.browaeys@iota.u-psud.fr).
\end{addendum}

\begin{methods}

\subsection{Derivation of the experimental signal.}
If $f_k({\bf r}) {\cal E}_k(t)$ is the field emitted by the atom $k$ ($k
= 1,2$), reference~\cite{Legero03} shows that the probability to detect
one photon at ${\bf r}_A$ and to detect the other one at ${\bf r}_B$, in
the 50/50 beam splitter configuration, after a delay $\tau$ is
proportional to
$$
w^{(2)}(\tau,{\bf r}_A,{\bf r}_B) =
\int |f_1({\bf r}_B) f_2({\bf r_A}) {\cal E}_1(t+\tau){\cal E}_2(t) -
f_2({\bf r}_B)f_1({\bf r}_A){\cal E}_2(t+\tau){\cal E}_1(t)|^2\, dt,
$$
which can be understood as the interference of two paths. Assuming a
temporal form of the field ${\cal E }_k(t) = H(t)\; e^{-\frac{\Gamma}{2}
t}\, e^{i\omega_{k}t}$ where $H(t)$ is the step function, we obtain
$$
w^{(2)}(\tau) \propto e^{-\Gamma |\tau|} (1 - K^2 \cos{\Delta \omega\; \tau}),
$$
where $K = |\int\, d{\bf r} \, f_1^*({\bf r})f_2({\bf r})|/\sqrt{
\int\, d{\bf r} \, |f_1({\bf r})|^2 \times \int\, d{\bf r} \,
|f_2({\bf r})|^2}$ is the spatial overlap of the electric field,
and $\Delta \omega$ is the frequency difference between the two
emitted photons. The double detection probability for $\tau=0$ is
proportional to $(1 - K^2)$, and so is the normalised ratio $R$
defined in the text. The proportionality factor is determined in
the absence of interferences ($K=0$): in this case, the two
photons behave as distinguishable particles and have a probability
of 1/2 to leave the 50/50 beam splitter through two different
ports. Thus the normalised ratio $R$ is $(1 - K^2)/2$.

\subsection{Model including the finite temperature of the atoms in the
trap.} To take into account the finite temperature of the atoms in
the trap, we integrate the expected signal for two photons
interfering with a frequency difference $\Delta \omega$ and a
spatial overlap $K$ (see section above), over the probability
distribution of frequency differences. To obtain this probability
distribution, we solve the equations of motion for a thermal
ensemble of single atoms in the trap, experiencing pulsed
excitations during 115\,$\mu$s, followed by a decay in a random
direction. After each pulse, we calculate the lightshifts of all
the atoms in the ensemble. We repeat this for 575 pulses to obtain
the distribution of lightshifts. This distribution is found to be
well represented by a function of the form $U^2
e^{-\frac{2U}{k_{\rm B} T}}$. The value of $\Delta \omega$ is
proportional to the difference in lightshifts experienced by the
atoms when they emit the photons. We then calculate the
auto-correlation of the lightshift distribution to get the
probability distribution of $\Delta \omega$. By averaging over
this distribution of lightshift differences, we obtain the
normalized coincidence rate signal as an analytical function with
only two fitting parameters, the spatial overlap and the
temperature of the atoms.

\end{methods}

\clearpage
\begin{figure}
\centering
\includegraphics[width=10cm]{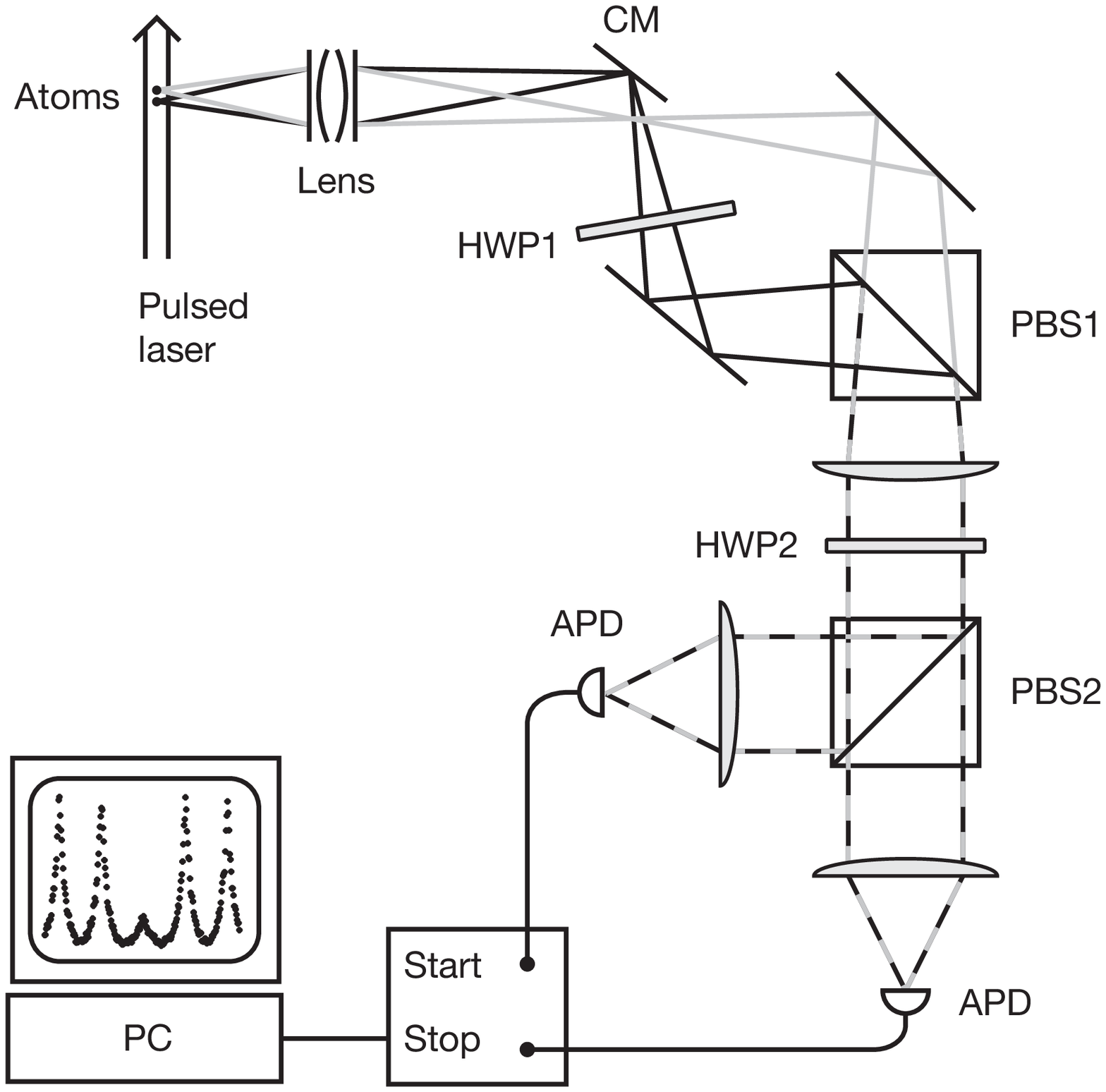}
\caption{Experimental setup. The two atoms are trapped in two
dipole traps separated by 6\,$\mu$m, and they are excited by the
same pulsed laser beam. The trap depth is 1.5\,mK, and the trap
frequency along the axis of the pulsed laser beam is 120\,kHz. The
emitted photons are collected by the same lens that is used to
create the dipole traps. The light from one of the traps is
separated off using a cut mirror (CM) placed close to the image
plane of the objective. In the plane of the cut mirror, the spot
corresponding to each trap has a waist of $\sim 90$ $\mu$m, and
the two images are separated by 500 $\mu$m. The half wave plate
HWP1 is oriented such that, at PBS1, the light beams from the two
atoms are recombined into the same spatial mode, but with
orthogonal polarizations. There are then two configurations to
detect the photons: either the axis of the half-wave plate HWP2 is
set so that the two orthogonal incident polarisations are equally
mixed in each output of PBS2, as in a 50/50 beam splitter, or the
axis is set so that the polarisations are unchanged, and then PBS2
simply separate the two beams coming from the two atoms without
mixing them. Two avalanche photodiodes (APD) are placed in the two
output ports of PBS2. The measured overall collection and
detection efficiency is 0.6\% for each photodiode~\cite{Darquie}.}
\label{figure1}
\end{figure}
\begin{figure}
\centering
\includegraphics[width=10cm]{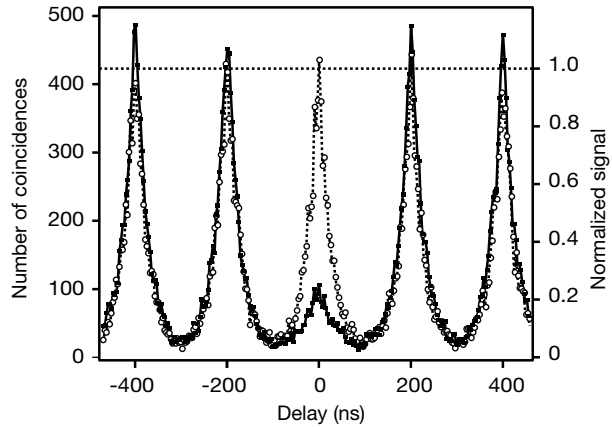}
\caption{Histograms of the time delays of the arrival of two
photons on the avalanche photodiodes, in the start-stop
configuration. Black squares correspond to the 50/50 beam splitter
configuration (interfering beams). Empty circles correspond to the
beam separator configuration (independent beams). These histograms
have been binned 3 times leading to a 3.6 ns resolution. The total
accumulation time is about 5 hours, corresponding to about 6600
events with two photons arriving on the beam splitter around zero
delay. The solid and dashed lines are a guide to the eye. The
normalized signal is obtained by dividing the number of counts by
the average value of the peak height in the beam separator
configuration.} \label{figure3}
\end{figure}
\begin{figure}
\centering
\includegraphics[width=10cm]{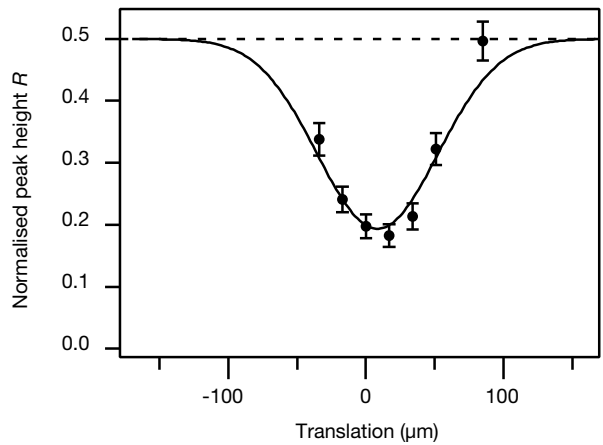}
\caption{Ratio of the height of the residual peak at zero delay in
the beamsplitter configuration to the average height of the peaks
in the beam separator configuration, as a function of the relative
distance between the two beams, translated parallel to each other.
The solid curve is the expected ratio, calculated from the
measured beam waist of the two beams. The amplitude and the center
of this curve is adjusted to fit the data. The error bars
correspond to statistical uncertainties.} \label{figure5}
\end{figure}
\begin{figure}
\centering
\includegraphics[width=10cm]{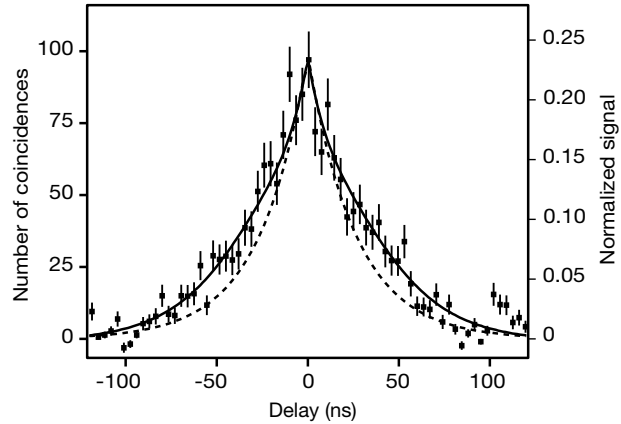}
\caption{Zoom of the histogram of Figure~2 in the 50/50 beam splitter
configuration, around zero delay. This curve is obtained from Figure~2
by subtraction of the contribution from the background and neighbouring
peaks. The squares represent the experimental data expressed in number
of coincidences. The solid line is a fit by the model described in the
Methods section, taking into account the finite temperature of the atoms
and the spatial overlap. The dotted line is the expected signal for zero
temperature. The error bars correspond to statistical photon counting
noise.}
\label{figure4}
\end{figure}
\clearpage
{\Large Supplementary Material}\\
The following appears as Supplementary Online Material in the published version.
\section*{A: THE NORMALIZED HEIGHT OF THE RESIDUAL PEAK FOR NON-INTERFERING PHOTONS}

In the absence of interference, i.e. if the spatial overlap $K=0$,
the height of the peak at zero delay after the normalization
described in the text is 0.5. The reason why it is 0.5 and not 1
can be understood from the following argument. The normalised peak
height, as described in the text, is simply the height of the peak
in the beam splitter configuration divided by the height of the
peak in the beam separator configuration, both taken at zero
delay. At zero delay, a coincidence event, whether in the beam
splitter or beam separator configuration, always corresponds to
the detection of one photon from each atom.  This is because both
atoms are near perfect single photon sources, and therefore the
probability that one of the atoms emits two photons during the
same excitation/emission cycle, which would appears as a peak et
zero delay, is negligible.

The difference between the two configurations is that in the beam
separator configuration, where each photodiode sees the light from
only one of the atoms, a pair of photons where one photon comes
from each atom always gives rise to a coincidence. In the beam
splitter configuration, both of the photons can end up at the same
photodiode, as each photodiode sees both atoms. In this case, no
coincidence occurs. For a 50-50 beamsplitter and distinguishable,
non-interfering photons, this happens  50\% of the time. The same
number of incident photon pairs therefore gives rise to half as
many concidences in this configuration and thus the ratio of the
heights of the peak at zero delay in the two configurations is
0.5.

\section*{B: ALIGNMENT OF THE OPTICAL SYSTEM AND LIMITS ON SPATIAL OVERLAP.}

In order to overlap the spatial modes of the two single photons in
free space, we used the fluorescence signal  of each of the two
single atoms induced by the magneto-optical trap laser beams.  We
measured the beam positions and waist ($1/e^2$ radius) in two
perpendicular directions and at two positions along the
propagation axis by taking intensity profiles using razor blades.
Using such profiles, the angular and translational alignment were
corrected step-by-step. This process was ultimately limited by the
error bars introduced by intensity fluctuations on the single atom
signal. The translational alignment (x and y) of the two spatial
modes was further improved using the contrast of the two-photon
interference signal itself, as shown in figure 4.

To understand how possible alignment errors contribute to the
spatial overlap, we have estimated how much the overlap changes in
the following situations:

\begin{enumerate}
\item The two beams have a different waist (different divergence).

\item The two beams are displaced transversally (x and y).

\item The position of the focal plane along the optical axis is
different for the two beams.

\item The two propagation axes have a small angle between them.
\end{enumerate}

In order to get an electric field mode overlap $K > 0.8$, one
should achieve better than a 4\% error on each of these
alignements, assuming that they are independent. As an example, if
the size of the waist of the two beams is different by 16\%, which
corresponds to our error bar on the waist size, then the overlap K
is already multiplied by 0.97. The cumulative effect of small
alignment errors seems therefore a reasonable explanation for the
limited spatial overlap that we observe. However, it should be
noted that phase errors across the wavefronts of the two beams due
to aberrations of the optical system would also decrease the
spatial overlap.

\section*{C: THE EFFECT OF INHOMOGENEOUS BROADENING ON THE SHAPE OF THE RESIDUAL PEAK AT ZERO DELAY}

The finite temperature of the atoms in the dipole traps gives rise
to an inhomogeneous broadening of the spectrum of the photons
emitted by the atoms. As described in our article, this broadening
manifests itself in our two-photon interference signal as an
increase in the width of the residual peak at zero delay. The
height of the residual peak, is not changed, and depends only on
the spatial overlap of the two beams.

In the case of perfect spatial overlap, the peak would disappear
at exactly zero delay. This is because if one looks close enough
around zero delay, the wavepackets of the two photons look alike
as dephasing due to their frequency difference has not had time to
occur. In a sense, perfect two-photon interference always occurs,
provided we look at ``short enough''  timescales. This is
equivalent to imposing temporal coherence by adding a narrow band
filter. In the case of perfect spatial overlap, our residual peak
would have a ``dip'' at exactly zero delay, with the width of this
dip depending on the inhomogeneous spectral width of the emitted
photons. This effect has been observed for two photons emitted by
the same source [4], and the theory is detailled in ref [22]. The
figure below summarizes this different configurations.
\begin{figure}
\centering
\includegraphics[width=10cm]{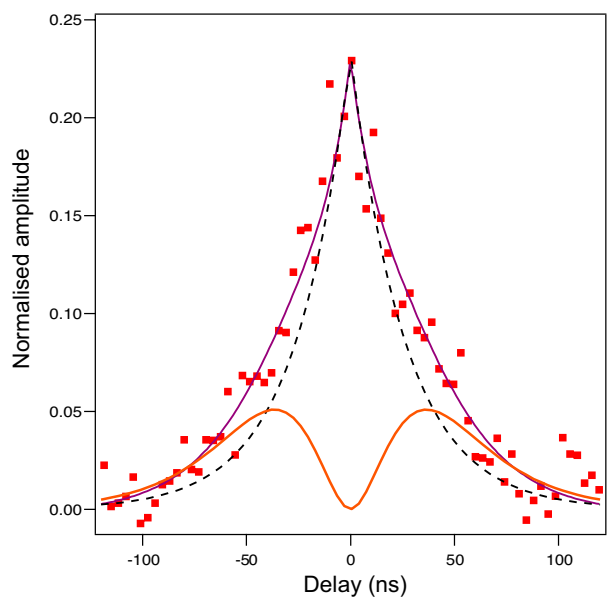}
\caption{Zoom of the histogram of Figure~2 in the 50/50 beam
splitter configuration, around zero delay. This curve is obtained
from Figure~2 by subtraction of the contribution from the
background and neighbouring peaks. The squares represent the
experimental data expressed in number of coincidences. The solid
line is a fit by the model described in the Methods section,
taking into account the finite temperature of the atoms and the
spatial overlap. The dotted line is the expected signal for zero
temperature. The solid line with a dip at zero delay is the
expected signal for a perfect spatial overlap and a temperature of
200 $\mu$K. This shows that at zero delay, in the case of a
perfect spatial overlap, the interference is always perfect,
whatever the temperature is.}
\end{figure}
\end{document}